# Adaptive Matching Pursuit based Online Identification and Control Scheme For Nonlinear Systems

Hamid Khodabandehlou

Electrical and Biomedical Engineering Department, University of Nevada, Reno
Reno, NV, USA
*e-mail:*hkhodabandehlou@nevada.unr.edu

Abstract: Complexity of adaptive control of nonlinear time varying systems requires the use of novel methods that have lower computational complexity as well as insuring good performance under time varying parameter changes. In this study, we use adaptive matching pursuit algorithm with wavelet bases for an online identification and control of nonlinear system with time varying parameters. We apply the proposed online identification and control scheme to two different benchmark examples of nonlinear system identification and control. Simulation results show that the proposed algorithm, using adaptive matching pursuit with wavelet bases, can effectively identify and control the nonlinear system even in presence of time varying parameters.

## 1. INTRODUCTION

Allgöwer and Zheng (2000) argued that despite availability of various adaptive control methods, application of these methods requires the knowledge and accessibility to an accurate mathematical model of the system. However, there are often un-modeled dynamics in most of real-world applications that raises serious questions on the performance of model-based control systems. As such any attempt using model free methods that have acceptable performance for controlling nonlinear systems are considered to be of great importance.

Findeisen and Allgöwer (2000) noted that while there are different model-based methods for adaptive control of linear systems, any novel approach in dealing with nonlinear systems requires also the use of suitable adaptive control method to insure satisfactory performance in such systems

Artificial neural networks have proven to be able to approximate any nonlinear system with desired accuracy hence neural networks received great deal of attention in identification of nonlinear systems (Khdabandehlou and Fadali, 2017). Neural network was used by Han et al. (2005) as an identifier for nonlinear system and using the model obtained by neural network; generalized predictive controller is designed and applied to nonlinear system.

However, despite the advantages gained by neural networks in nonlinear system identification, their slow convergence rate and computational complexity raises serious questions on their application for an online identification and control of nonlinear systems. To resolve this problem, control engineers have proposed the use of wavelet network where the structure of this network is completely similar to neural networks and where the only difference is in activation functions. In wavelet networks activation functions are wavelets which may have different scales and shifts. Q.zhang and Beneveniste (1992) showed that wavelet networks can approximate any nonlinear function the same as done by neural networks.

They also argued that for given nonlinear system and desired accuracy of approximation, wavelet neural network may have fewer nodes as compared with artificial neural networks and hence wavelet neural networks may be a better choice for an online identification and control of nonlinear systems.

Sousa et al. (2002) used a wavelet network as a model identifier for identification and control of robot and the results are compared with the case of using neural network. Stability of the closed loop system is insured based on the second method of Lyapunov.

Zayeni and Ahmadi applied a Radial wavelet network for identification of nonlinear system. Structure of this network is similar to structure of Radial basis function networks in which learning method used is also similar to this network where the only difference is that nodes activation functions are wavelets.

Self-Recurrent wavelet neural network with adaptive learning rate was used as a model identifier by Yoo et al. (2006). Based on this model, generalized predictive controller is designed for nonlinear system and where the stability of the close loop system is proved using the Lyapunov method.

Wavelets have been widely used in identification and control of linear and nonlinear systems. Khodabandehlou et. al. (2018) used wavelet neural network and model predictive controller to control a seismically isolated structure during earthquake. Their simulation results show that wavelet neural network based controller can effectively control the structure during near fault and far field ground motions. Khodabandehlou and Fadali (2017) used wavelet neural network with feedforward component to control an unmanned vehicle over the communication channel. Their simulation results show that wavelet neural network with feedforward component can effectively identify the model of the unmanned vehicle in presence of fixed and random network delays and packet loss.

Shmilovici and Maimon applied adaptive matching pursuit algorithm to identification of nonlinear system and using this method with Spline bases, they identified and controlled nonlinear system. Algorithm is applied to several nonlinear systems in which simulation results show that algorithm yields high performance in identification of nonlinear systems. It is shown that due to localization property of wavelets during approximation, algorithm has low computational complexity and where with current devices can be implemented easily.

In this work we use adaptive matching pursuit with wavelet bases for identification and control of nonlinear time varying system. We apply the algorithm to two type of nonlinear time varying systems: systems with slow parameter changes and systems with fast changes in parameters. Simulation results show that wavelet bases can lead to high performance in identification and control of the given process where changes in parameters in both cases poses no major difficulty on the tracking of the closed loop system. It is shown that the changes in parameters may have direct effect on the closed loop input where based on the case we study, they may cause the control input to become smaller or great.

Sections of this paper are organized as follows. Section 2 describes the matching pursuit algorithm as proposed by S.G.Mallat. Section 3 describes the adaptive matching pursuit algorithm and section 4 describes the application of adaptive matching pursuit to adaptive control followed by the illustration of simulation results shown in section 5.

## 2. MATCHING PURSUIT

The matching pursuit proposed by S.G.mallat solves the following problem:

Given a collection of vectors $D = \{g_\gamma\}_{\gamma \in \Gamma}$ in the Hilbert space where all the vectors have unity norm, it is desired to describe a given function $f$ using these bases. This method is similar to the Projection Pursuit that is used in Statistics. Collection of vectors D is referred to as dictionary where vectors are called atoms analogous to atoms as basic entities of the given ensemble. The problem of vector/function expansion using D can be formulated as follows:

Assume that $D = \{g_\gamma\}_{\gamma \in \Gamma}$ be a dictionary with $P > N$

Atoms of unity norm which belong to the Hilbert space. This dictionary has $N$ linearly independent vectors that form a basis for $C^N$. A representation of the form of eq.(1) for function $f$ is to be found where $P_v$ indicates the orthogonal projection on space $V = \text{span}\{g_n\}$.

$$P_v f = \sum_n a_n g_n \quad (1)$$

The Algorithm begins with projection of $f$ onto $g_{\gamma 0} \in D$ and calculation of residue $Rf$:

$$f = <f, g_{\gamma 0}> g_{\gamma 0} + Rf \quad (2)$$

And hence $Rf$ is orthogonal to $g_{\gamma 0}$:

$$\|f\|^2 = |<f, g_{\gamma 0}>|^2 + \|Rf\|^2 \quad (3)$$

For minimizing $\|Rf\|$, the atom $g_{\gamma 0} \in D$ have to be chosen such that $|<f, g_{\gamma 0}>|$ is maximum. In the next iteration, $|<Rf, g_{\gamma 1}>|$ is chosen to be maximum and algorithm continues the same way. In the M'th iteration of the algorithm, an intermediate representation of function $f$ can be expressed as follows:

$$f = \sum_{m=0}^{M-1} <R^m f, g_{\gamma m}> g_{\gamma m} + R^M f = f^k + R^k f \quad (4)$$

In general, in each iteration of the algorithm, the following operations have carried-out.:

1) Calculate the projection of $R^k f$ on all of the dictionary elements.
2) Find the index $\gamma_{k+1}$ for which the projection is maximal.

$$\sup_{\gamma_{k+1}} |<R^k f, g_{\gamma_{k+1}}>| \quad (5)$$

3) Update the model

## 3. IDENTIFICATION WITH ADAPTIVE MATCHING PURSUIT

Schmilovici and Maimon argued that under the condition where the signal to noise ratio is high enough and the identification process is sufficiently faster than the changes of system dynamics, then the subspace at any given time can be approximated as a single point of $f \in H$. So at any instance, it is sufficient to modify only one of the model coefficients that produce the largest residual error and the other coefficients will remain unchanged.

Implementation of the algorithm in finite vector space depends on the definition of the nonlinear function to be identified. In this work we will consider the model as a nonlinear auto-regressive with exogenous input (NARX) as in (6).

$$y(k) = f(y(k-1), ..., y(k-p); u(k-1), ..., u(k-q)) + n(k) \quad (6)$$

Where $f$ is the function to be identified and $u, y \in R$ are system input and output respectively and $p, q$ are positive integers and $n(k)$ is the measurement noise which is assumed to be white (due to NARX formulation). Define:

$$\bar{\varphi} = [\varphi_1(k), ..., \varphi_n(k)] = [y(k-1), ..., y(k-p); u(k-1), ..., u(k-q)] \quad (7)$$

In general, we want to approximate a function of measurements in the form of basis expansion as given in the following form.

$$\hat{f}(\overline{\varphi}(k)) = \hat{f}(k) = \overline{g}_\varphi^T(k)\theta(k-1) \qquad (8)$$

Where $\overline{\theta}(k)$ is an n dimensional vector that contains the model coefficients at iteration k and therefore we have:

$$\overline{g}_\varphi(k) = [g_1(\overline{\varphi}(k)),...,g_2(\overline{\varphi}(k))] \qquad (9)$$

Where $g_i(\overline{\varphi}(k))$ represents the projection of vector $\overline{\varphi}(k)$ on the basis $g_i$. For an online implementation of the algorithm, a process in the form of (10) is to be found to update $\overline{\theta}$ to follow the changes of $f$.

$$\overline{\theta}(k) = \overline{\theta}(k-1) + \overline{w}(k) \qquad (10)$$

Where $\overline{w}(k)$ represents the changes in the parameters vector due to approximation and adaptation errors. With this assumption, the following procedure is proposed for updating $\overline{\theta}$:

1) Calculate $\overline{g}_\varphi(k)$
2) Find index $m(k) \in [1,...,n]$ for which the element $|g_{m(k)}(\overline{\varphi}(k))|$ in the vector $|\overline{g}_\varphi(k)|$ is maximum.
3) Update the model.

Define:
$$e(k-1) = y(k) - \overline{g}_\varphi^T(k)\theta(k-1) \qquad (11)$$

Consequently, the equation according to which $\theta$ is updated will be as (12).

$$\overline{\theta}(k) = [\theta_1(k-1),..., \\ \theta_{m(k)-1}(k-1), \theta_{m(k)}(k-1) + \\ \frac{e(k-1)}{g_{m(k)}(\overline{\varphi}(k))}, \theta_{m(k)+1},...,\theta_n(k)] \qquad (12)$$

Due to point wise estimation, the sensitivity of the algorithm to the number of basis functions is not too high. However in practical implementations, since the number of basic functions has great effect on the computational complexity, convergence rate of the algorithm and the error covariance, selection of suitable basis functions is of great importance. Simulation results show that if the number of basis functions is lower than $n$, the approximation error may increase rapidly which will lead to high tracking error which in turn leads to output distortion which will be explained in next section. In order to avoid such problems, the conservative choice would be to choose at least $n$ basis functions.

The convergence of algorithm, as in other adaptive methods, depends on the statistical properties of the algorithm inputs. Shmilovici and Maimon proposed a proof to algorithm with certain assumptions on the algorithm inputs. These assumptions are too conservative and are usually violated in real world applications. Using lemma proposed by Ljung and Priouret (2007), the assumptions can be relaxed so that the convergence of the algorithm can be insured. The proposed method under relaxed assumptions is still convergent and also able to identify the nonlinear system but the convergence limit is different and the identification error is greater than the previous case.

## 4. APPLICATION TO ADAPTIVE CONTROL

In this paper the nonlinear time varying system is considered as (13).

$$y(k) = f(y(k-1),...,y(k-p); \\ u(k-1),...,u(k-q)) + u(k) \qquad (13)$$

Where $f$ is the nonlinear system to be identified and $u, y \in R$ are system input and output respectively and the vector $[y(k-1),...,y(k-p);u(k-1),...,u(k-q)]$ is assumed to be measurable. The control objective is that the system output follows the output of the n'th order linear stable system as (14).

$$y_m(k) = \sum_{i=1}^n s_i y_m(k-i) + r(k) \qquad (14)$$

Where $\{s_i\}_{1 \leq i \leq n}$ are the coefficients of the reference model characteristic equation and $r(k)$ is the reference input. With this assumption, the reference input will go through the low pass filter which smoothens the reference signal that in turn will lead to smoothness of the closed loop input and smaller closed loop input because sharp changes in reference signal may force the controller to imply great controlling input. The $\{s_i\}_{1 \leq i \leq n}$ should be chosen so that all of the desired roots of the characteristic equation $h(z) = z^n + s_1 z^{n-1} + \cdots + s_n$ lie in the unit circle so that the desired model be stable and well behaved.

The error between the system output and the reference model output can be defined as $e(k) = y_m(k) - y(k)$. The nonlinear system approximation error can be defined as $\eta(k) = f(k) - \hat{f}(k)$, where $\hat{f}(k)$ is the approximation of the nonlinear function $f(k)$ at iteration $k$. $\eta(k)$ is referred to as matching condition. The objective of the adaptation algorithm objective is to insure asymptotic tracking as given by (15) as follows.

$$e(k) + s_1 e(k-1) + \cdots + s_n = 0 \qquad (15)$$

In the case of prefect matching condition, $\eta(k) = 0$, therefore it is expected that $\lim_{k \to \infty} e(k) = 0$ which can be which can be achieved by the control law as given below.

$$u(k) = [-f + \overline{s}^T \overline{\varphi}(k) + r(k)] \qquad (16)$$

By substituting (16) in (13) and subtracting the result from (14),the (15) will be obtained. However, since the nonlinear function is unknown, in (16) $f$ will be replaced with $\hat{f}_k$ obtained from (8). Therefore the control law will be as (17).

$$u_c(k) = [-\hat{f}_k + \bar{s}^T \bar{\varphi}(k) + r(k)] \quad (17)$$

This equation is called the certainty equivalence controller which implies that in the case of prefect matching, this controller will be equivalent to the controller that is designed if the nonlinear function was known.

In the previous section it was mentioned that if the number of basis functions be lower than $n$ will leads to output distortion. From (13) and (17) it is obvious that $y(k) = y_m(k) + \eta(k)$ and from this equation it can be realized that the identification error directly affects the system output. The general scheme of identification and control is shown in the Fig.1

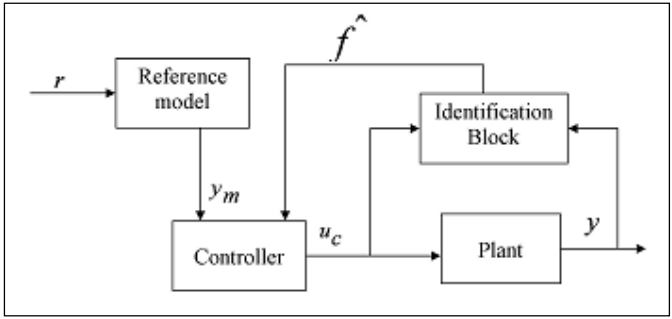

Fig.1 : general scheme of identification and control

## 5. SIMULATION RESULTS

As an illustration of the application of the proposed method, we apply the algorithm to two sets of systems: 1) system with slow parameter changes and 2) system with sufficiently fast parameter changes. The systems are chosen based on the systems used by shmilovici,Maimon and Zayeni, Ahmadi with some modifications. In order to have nonlinear time varying behavior, time varying coefficients are considered in the models. The sensitivity of the algorithm to the number of basis functions is not too high, but we have chosen at least $n$ basis functions to ensure a desirable identification performance. Also the algorithm is indifferent to the type of basis functions but for good identification and tracking performance, it is preferable to choose the basis functions that have similar behavior as the given nonlinear function to be identified which can be done by having a set of input-output data of the system. In this work, depending on the nonlinear function, DB2, DB3, DB3, DB4, DB5, BIOR3.1, BIOR3.3 wavelets and their corresponding scaling functions are used as basis functions. Since the values of the considered wavelets and scaling functions tend to zero around $t = 10s$, basis functions were shifted to fully cover this region.

In recursive methods, one of the convergence conditions is that the input signal to be enough exciting. In this case it can be concluded that the algorithm converges rapidly, if $y_m$ be enough exciting.

Also it should be mentioned that the numerical estimation of this wavelets and scaling functions have been calculated using MATLAB. Due to point wise estimation of the algorithm, the simulation time is fast enough and the algorithm can be implemented in on line identification and control applications.

### 5.1 Example 1

In this example the nonlinear time varying system is expressed as follows:

$$y(k) = f[y(k-1), y(k-2), u(k-1)] + u(k) + n(k) \quad (18)$$

Where

$$f[y(k-1), y(k-2), u(k-1)] = \frac{y(k-1)[a(k)u(k-1)y(k-2) + 2.5]}{1 + y(k-1)^2 + y(k-2)^2 + u(k-1)^2}$$

And $a(k)$ is time varying coefficient whose value is initially one and increased 20% from $t = 0s$ to $t = 25s$. According to (18) at least three basis functions are required. The characteristic equation of the reference model whose output is to be followed by the system is considered of order three and is expressed as (19).

$$y_m(k) = 1.3 y_m(k-1) - 0.72 y_m(k-2) + 0.16 y_m(k-3) + r(k) \quad (19)$$

Where $r(k)$ is the reference input. Reference model has three poles in unit circle which are placed at $0.5$ and $0.4 \pm 0.4i$ .using (16) the closed loop input can be expressed as (20).

$$u_c(k) = -\hat{f}(k) + 1.3 y_m(k-1) - 0.72 y_m(k-2) + 0.16 y_m(k-3) + r(k) \quad (20)$$

Where the reference input is sinusoidal and the measurement noise, $n(k)$, is assumed to be white Gaussian noise with standard deviation of 0.01. Basis functions are defined in the region $[0,10]$ seconds and assumed to be periodic (Fig.2).
In Fig.3 the system output and the reference model output are shown where Fig.4 depicts the closed loop input.

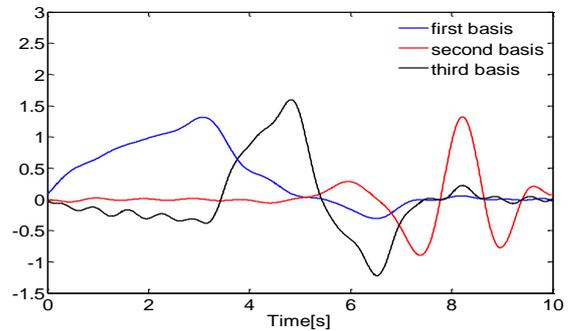

Fig.2. Basis function of example 1

In $t = 50s$ it is observed that the error increases substantially(Fig.2) which can be explained according to (12) in which $g_{m(k)}(\bar{\varphi}(k))$ tends to zero in some periods of time which in turn will be problematic because of division to zero in (12). The intuitive idea would be to consider a very small value ($\varepsilon$) in the denominator of (12) in such cases. However this choice seems to be suitable, but this small value tends to undesirable increase of one of the coefficients. This can result in an unacceptable error for long time. Hence it is preferred that the value of $\varepsilon$ not to be too small. However large values for $\varepsilon$ will lead to the output similar to the one shown in the fig.3.

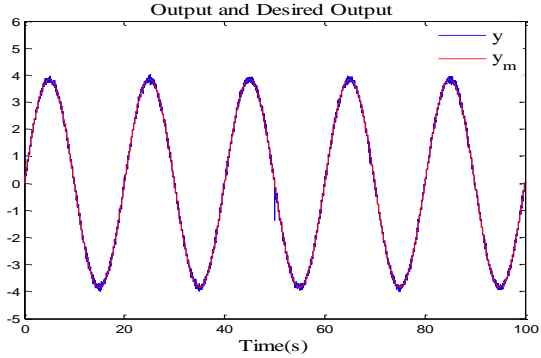

Fig.3. System output and the reference model output

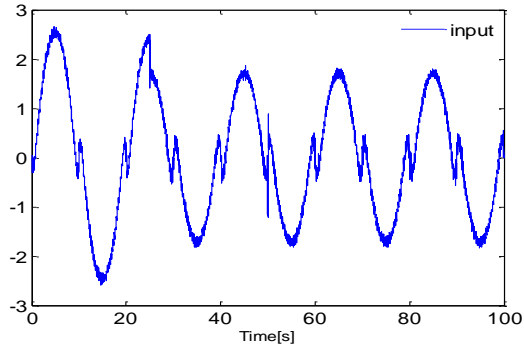

Fig.4. Closed loop input

*5.2 Example 2*

The nonlinear system of this example is based on [6] which is also modified to have time varying behaviour. The nonlinear system is as follows:

$$y(k) = (0.8 - 0.5e^{-y(k-1)^2})y(k-1)a(k)$$
$$- (0.3 + 0.9e^{-y(k-1)^2})y(k-2) \quad (21)$$
$$+ 0.1\sin(\pi y(k-1)) + u(k-1) + n(k)$$

Where $a(k)$ is time varying coefficient whose value is initially one and changes to three at $t = 25s$ and $n(k)$ is the measurement noise which is assumed to be white Gaussian noise with standard deviation of 0.01. In spite of this change, the controller performance is acceptable and the effect of this variation on the closed loop input is noticeable. The reference model poles are placed at $0.4$ and $0.2 \pm 0.2i$ Using (16), the closed loop input will be as given by (22).

$$u_c(k) = -\hat{f}(k) + 0.8y_m(k-1) -$$
$$0.24y_m(k-2) + 0.032y_m(k-3) \quad (22)$$
$$- u(k-1) + r(k)$$

In this example tracking error and control input depend on the choice of the poles of reference model and as the reference model poles get closer to the origin, the controller performance in tracking improves whereas closed loop input tends to be oscillating which is undesirable hence there should be a trade-off between tracing error and the closed loop input smoothness.

Fig.5 shows the system output and the reference model output whereas Fig.6 and Fig.7 are showing the closed loop input and basis function respectively.

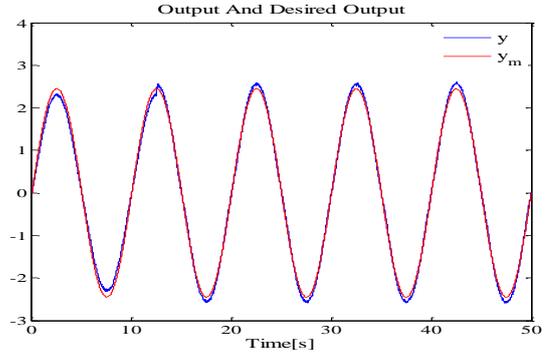

Fig.5. System output and the reference model output

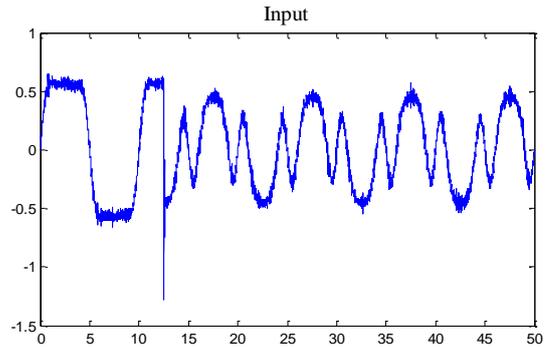

Fig.6. Closed loop input

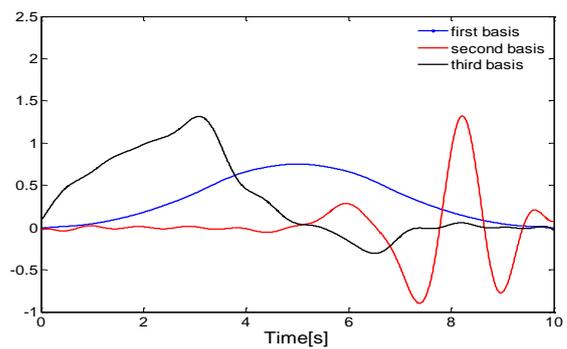

Fig.7. Basis function of example 2

From fig.4 and fig.7 it can be concluded that the measurement noise has direct effect on the closed loop input and will make it to be oscillating.

## 6. CONCLUSION

In this paper, adaptive matching pursuit algorithm with wavelet bases was used for an on-line identification and control of nonlinear time varying system. Time varying coefficients are set to slow and fast variations. Simulation results show that controller performance is acceptable even if model coefficients are time varying and have high variations as of example 2. it was also shown that the variation of parameters will affect the closed loop input, and depending on the manner these parameters are changed as well as the given nonlinear system, a change is observed in the closed loop input and base on the nonlinear system, the control input may became smaller or greater. In the case of fast varying coefficients, a great identification error occurs which in turn will lead to great change in control input.